\documentclass[aps,twocolumn,tightenlines,letterpaper]{revtex4}

\usepackage{amsmath}
\usepackage{graphicx}

\usepackage{amsfonts,amssymb,amsmath,amsthm}
\usepackage{hyperref}
\usepackage{graphicx}

\newcommand{\be}{\begin{equation}}
\newcommand{\ee}{\end{equation}}
\newcommand{\ba}{\begin{array}}
\newcommand{\ea}{\end{array}}

\newcommand{\calH}{{\cal H }}
\newcommand{\calL}{{\cal L }}
\newcommand{\calM}{{\cal M }}

\newcommand{\calZ}{{\cal Z }}
\newcommand{\calS}{{\cal S }}

\newcommand{\CC}{\mathbb{C}}
\newcommand{\la}{\langle}
\newcommand{\ra}{\rangle}

\newtheorem*{theorem}{Theorem}

\newtheorem*{lemma}{Lemma}

\newcommand{\ncd}{\newcommand}
\ncd{\QCc}{$QC_{\cal{C}}\;$}
\ncd{\QCcns}{$QC_{\cal{C}}$}

\begin{document}

%{\it \hspace{15cm} \large \it DRAFT}

\title{On measurement-based quantum computation with the toric code states}
\author{Sergey Bravyi${}^1$ and Robert Raussendorf${}^2$}
\affiliation{${}^1$\mbox{IBM Watson Research Center, Yorktown Heights, NY 10598 USA} \\
${}^2$\mbox{\mbox{Perimeter Institute, Waterloo, Canada N2L 2Y5}}}

\date{\today}

\begin{abstract}
We study measurement-based quantum computation (MQC) using as quantum
resource the
planar code state on a two-dimensional square lattice
(planar analogue of the toric code).
It is shown that MQC with the planar code state can be
efficiently simulated on a classical computer if
at each step of MQC the sets of measured and unmeasured qubits
correspond to connected subsets of the lattice.
\end{abstract}

\maketitle

\section{Introduction}

Quantum mechanical systems allow, for a class of computational
problems, exponentially more efficient processing of information than classical
systems. The question of where this speedup comes from has been under
intensive debate over the recent years. `Largeness of
Hilbert space' \cite{Fey}, `entanglement' (see e.g. \cite{Vid}) and
`superposition and interference' \cite{QArv}, for
example, have been suggested. They all seem to be a part of the puzzle
but it is difficult to pin
down a single characteristic property.

A prerequisite for a quantum speed-up in computation with a
quantum system is the hardness of its classical simulation. One
may learn about the cause for the quantum speed-up by
investigating the circumstances under which it vanishes, i.e. when
efficient classical simulation becomes available. A number of such
scenarios have been described in the literature. For example, the
unitary evolution of a spin chain can be followed efficiently as
long as the entanglement with respect to all possible
bi-partitions remains small \cite{Vid}. Further, linear optics
with Gaussian states, systems of non-interacting fermions and
qubit registers acted upon by gates from the so-called Clifford
group are quantum systems which can be efficiently simulated
classically; see \cite{Gied}, \cite{TeDiV,SB04} and \cite{Go},
respectively.

Here we address classical simulation of quantum systems in the
context of measurement-based quantum computation (MQC), in particular
the one-way quantum computer (\QCcns) \cite{RB01}. In this scheme,
one-qubit measurements are performed on a
multi-qubit entangled resource state, the so-called cluster state.
After the universal cluster state is created, no further
interaction among the qubits takes place. Quantum information is
written onto the cluster, processed and read-out from the cluster by
the one-qubit measurements alone.

To obtain a better understanding about MQC
one may apply certain alternations to the original scheme~\cite{RB01}
and search for properties which remain invariant under these
changes. Which other schemes for the processing of the measurement
outcomes exist? Which other quantum states are universal resources
and which properties characterize them? With regard to the classical
processing of
measurement outcomes in MQC, modified schemes have been described in
\cite{Danos, Hall, Eis}. Alternative universal resources for MQC have been
presented in \cite{Debbie, VdN1, Eis}. For example, universal resource
states exist in which each qubit is arbitrarily close to a pure
state \cite{Eis}. Concerning universal vs. efficiently simulatable
quantum resource states, a systematic study has begun in
\cite{Shi1, Shi2, VdN1, VandenNest06}. Diverging amounts of
entanglement, measured in terms of the so-called entanglement width,
are required for universality \cite{VdN1,VandenNest06}. On the
other side of the spectrum, MQC can be simulated efficiently
classically by a so-called tree tensor network (TTN) if
the resource state is a graph state and the graph is a tree or close
to a tree \cite{Shi1, Shi2, VandenNest06}. The prototypical example is the
graph state on a line graph, i.e., the 1D cluster state
\cite{Nielsen}. An example for large deviation from tree-ness is the
universal 2D cluster state whose TTN simulation is thus hard.

In this paper, we describe a complementary simulation method for MQC,
centered around
planarity of graphs. The counterpart of the 1D cluster state is the
planar code state \cite{Kit1, Kit2}, and that of the TTN is the
partition function of the Ising model. Within our framework, an
example for large deviation from planarity again is the 2D cluster
state.

Planar code states and cluster states are closely
related. For example, if
one applies a certain pattern of Pauli-measurements
to the two-dimensional cluster
state, one can prepare the planar code state \cite{RBH04}.
%(our results indicate that the reverse conversion is highly unlikely).
One dimension higher up, the fault-tolerance properties found in
three-dimensional
cluster states \cite{RBH04} are related to the `Random plaquette
    $\mathbb{Z}_2$ gauge model in three dimensions' which also describes
fault-tolerant data storage with a planar code \cite{DKLP}.

A further interesting property of the planar code state is that it
obeys the entanglement area law~\cite{Zanardi04}.
That is, the entanglement entropy of a block of spins is proportional to its
perimeter. Thus bi-partite entanglement in the planar code state is
large. This state also exhibits topological quantum order and it is therefore
not possible to prepare the planar code state from a product state by
a small-depth unitary quantum circuit~\cite{Bravyi06}.

However, our result is that MQC with the planar code state as the
quantum resource is not universal and can be simulated efficiently
classically. This unexpected property of the planar code state can be
attributed to the exact solvability of the Ising model on a planar
graph. Thus, although large entanglement
in the resource state is necessary for MQC, it is not sufficient.

\section{The planar code state and MQC}

\begin{figure}
\centerline{
\mbox{
 \includegraphics[height=4cm]{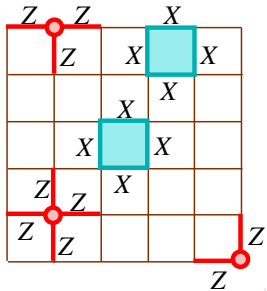}
 }}
\caption{ \label{fig:planar}Stabilizer generators for the planar code state
      $|K\ra$. Qubits are located on the edges of the lattice.
      $Z$-type stabilizer generators live on vertices, and
      $X$-type stabilizer generators live on plaquettes.}
\end{figure}

We consider two-dimensional square lattice of dimensions $L\times L$. It consists
of $N_2=L^2$ plaquettes, $N_1=2L(L+1)$ edges, and $N_0=(L+1)^2$ vertices.
Qubits live on the edges of the lattice, so the Hilbert space is
$\calH=(\CC^2)^{\otimes N_1}$.
Let $X_e$ and $Z_e$ be the Pauli operators $\sigma^x$ and $\sigma^z$ acting on the qubit of edge $e$
tensored with the identity operators on all other edges.
For any vertex $s$ and any  plaquette $p$
define {\it stabilizer generators}
\begin{equation}
  \label{SG}
  A_s = \prod_{e\in \delta s} Z_e,\quad  B_p =
  \prod_{e \in \partial p} X_e.
\end{equation}
Here $\delta s$ denotes the set of edges incident to vertex $s$, and
$\partial p$ denotes the set of edges making up the boundary of plaquette $p$,
see Figure~\ref{fig:planar}. Any pair of generators commute with each other since
the sets $\delta s$ and $\partial p$ always share even number of edges.
All $N_2$ plaquette-type generators  $B_p$ are independent. As for the vertex-type
generators $A_s$,
only $N_0-1$ of them are independent, since $\prod_{s=1}^{N_0} A_s = I$.
Thus the total number of independent generators is $N_0+N_2-1=2L(L+1)$.
It coincides with the number of qubits $N_1$, so there exists a unique
{\it planar code state} $|K\ra\in \calH$ satisfying the stabilizer
equations
\be\label{stabilizers}
A_s\, |K\ra=|K\ra, \quad  B_p\, |K\ra = |K\ra,
\ee
for all $s=1,\ldots,N_0$ and
$p=1,\ldots, N_2$.
It is worthwhile to write down the state $|K\ra$ explicitly.
We shall use the {\it computational basis} of $\calH$. Accordingly, a basis vector
$|x\ra$ represents a configuration of $N_1$ binary variables
$x=\{x_e\}_{e=1,\ldots,N_1}$,
$x_e\in \{0,1\}$
living at the edges  of the lattice.
We shall call such a configuration a {\it $1$-chain}.
 Let us call a $1$-chain $x$ a {\it $1$-cycle} iff
$\sum_{e\in \delta s} x_e$ is even for every vertex $s$.
In other words, a $1$-cycle must have even number of edges incident to any vertex.
Let $\calZ\subset \{0,1\}^{N_1}$ be a set of all $1$-cycles.
Then the  state $|K\ra$ can be explicitly written in the computational basis
as
\be
|K\ra= \frac1{\sqrt{|\calZ|}}\, \sum_{x\in \calZ} |x\ra.
\ee
One can easily check that $|\calZ|=2^{N_2}$, that is each plaquette contributes
one independent $1$-cycle to $\calZ$.

MQC with the planar code state  is defined as adaptive sequence of one-qubit
measurements $M_1,M_2,\ldots,M_{N_1}$
applied to the state $|K\ra$.
After all measurements are done, every qubit on the lattice is measured in some basis.
The choice of the measurement
$M_{j+1}$ may be determined by the outcomes of all earlier measurements
$M_1,\ldots,M_j$.
We allow only complete von~Neumann  measurements.
In other words, $M_j$ is specified by a unity decomposition
\[
I=|\psi_j^0\ra\la \psi_j^0| + |\psi_j^1\ra\la \psi_j^1|
\]
for some orthonormal basis $|\psi_j^0\ra, |\psi_j^1\ra\in \CC^2$.
Let $e_j$ denote the qubit to be measured at step $j$ and $m_j\in \{0,1\}$ be
the outcome of the measurement $M_j$.
After the measurement the qubit $e_j$ is projected onto a state $|\psi_j^{m_j}\ra$.
Thus MQC is specified by
\begin{itemize}
\item Description of the first measurement\\ $M_1=(e_1,|\psi_1^0\ra,|\psi_1^1\ra)$
\item Efficient algorithm that takes as input outcomes $m_1,\ldots,m_j$ of the first
$j$ measurements
and returns a description of the next measurement\\
$M_{j+1}=(e_{j+1},|\psi_{j+1}^0\ra,|\psi_{j+1}^1\ra)$.
\end{itemize}
The output of MQC is a string of outcomes $m_1,\ldots,m_{N_1}$
having some probability distribution $p(m_1,\ldots,m_{N_1})$.
We say that MQC can be efficiently simulated classically
if there exists
a classical randomized algorithm running in a time $poly(L)$ that
allows one to sample $m_1,\ldots,m_{N_1}$ from
the probability distribution $p(m_1,\ldots,m_{N_1})$.

In general, there are no restrictions on the order in which the qubits
are measured in MQC.  We shall consider more restrictive settings
(because our technique does not apply to the general case).
Consider a set $E_j=(e_1,e_2,\ldots,e_j)$ including all qubits
measured up to the step $j$ of MQC.
Let $\bar{E}_j$ be the complimentary set including all unmeasured qubits.
Let us say that a set of edges $E$ is {\it connected} iff
for any pair of edges
$e,e'\in E$ there exists a path $e_0=e,e_1,e_2,\ldots,e_n=e'$
that connects $e$ with $e'$ and such that $e_l\in E$ for all $l$.
Our main result is
\begin{theorem}
\label{thm:MQC}
Suppose at each step $j=1,\ldots,N_1$ of MQC the sets of measured
and unmeasured qubits $E_j$, $\bar{E}_j$ are connected.
Then
MQC can be efficiently simulated classically.
\end{theorem}
The proof relies on two observations. The first observation
made by Kitaev~\cite{Kitaev-pc} is that
the overlap problem --- computing the inner product between $|K\ra$ and an arbitrary product state
can be reduced to computing  partition function of the Ising model with
complex weights.
The connection between overlaps of states and partition functions of
models in statistical mechanics
is explored in greater generality in \cite{VdN2}.

More strictly, we use Barahona's techniques~\cite{Barahona81}
to express the overlap as a generating function of perfect matchings
on a properly defined planar graph. It was shown by
Kasteleyn~\cite{Kasteleyn61} that such generating functions can be computed efficiently,
see~\cite{Galluccio98} for more recent exposition.
Thus if every qubit on the lattice is
measured in some specified basis, probability of any particular outcome
can be computed efficiently.

However, this observation solves only part of the problem.
First of all, since the number of possible outcomes, $2^{N_1}$, is exponentially large,
computing probability of any particular outcome does not allow us to
{\it sample} the outcomes according to these probabilities.
More importantly, since the measurements in MQC are {\it adaptive}, we must be able to compute
probabilities of {\it partial} measurements, when only some subset of
qubits is measured. Our second observation is that the overlap problem for
a subset of qubits can be reduced to computing partition function of the Ising model
for two independent replicas of the lattice with properly identified boundaries.
The constraint that the subsets $E_j,\bar{E_j}$ are connected
is the simplest way to ensure that the doubled lattice corresponds to a planar graph.
The partition function of Ising model on any planar  graph
can be computed efficiently using Barahona's algorithm~\cite{Barahona81}.
Given probabilities of any partial measurement outcomes, we can compute
{\it conditional} probabilities and thus we can simulate MQC.

Whether or not the connectivity constraint is a severe restriction on MQC is an open question.
Note that this constraint is trivially satisfied in the standard scheme of MQC, when the
qubits are measured in the order ``from the left to the right".
We also conjecture that the theorem above can be generalized to the following scenarios:
(1) the number of connected components in $E_j,\bar{E}_j$ is greater than one,
but can be bounded by a constant. (2) The square lattice on a plane
can be replaced by a lattice on any two-dimensional
orientable surface with genus $g$ bounded by a constant.
The intuition supporting these conjectures comes from the
work~\cite{Galluccio98} providing
an efficient algorithm to compute the generating function of perfect matching
if the genus of the graph can be bounded by a constant.

\section{Classical simulation}
\label{CS}

This section is organized as follows.
Section~\ref{subs:o1} describes the reduction from the overlap problem to
the Ising model and then to the perfect matchings on a planar graph.
It mainly follows the reference~\cite{Barahona81} (except for considering
complex weights) and aims to make our exposition self-sufficient.
Section~\ref{subs:o2} shows how to describe a mixed state
obtained from $|K\ra$ by the partial trace over a subset of qubits
as a mixture of planar code states.
Finally, Section~\ref{subs:o3} explains how to compute probability of
any partial measurement outcome by taking two copies of the lattice.

\subsection{The overlap between planar code states and product states}
\label{subs:o1}

In this section we consider planar code states on subgraphs
of the square lattice~\footnote{Note that at each step of MQC only some part
of qubits is measured. To compute probabilities of outcomes of such
partial measurements
we have to consider arbitrary subgraphs of the square lattice.}.
We show that the overlap (the inner product) between the planar code
state and a product of any one-qubit states can be
efficiently computed using techniques developed by
Kasteleyn and Barahona~\cite{Kasteleyn61,Barahona81}.

Suppose the planar code state $|K\ra$ is defined on a square lattice $\calL$,
see Figure~\ref{fig:planar}.
Denote $V(\calL)$ and  $E(\calL)$ sets of vertices and  edges
of $\calL$. For any subset of edges $E\subseteq E(\calL)$ define a planar graph
$G_E=(V,E)$, where $V\subseteq V(\calL)$ is a set of all vertices $s\in V(\calL)$
having at least one incident edge from $E$,
\[
V=\{ s\in V(\calL)\, : \, \delta s \cap E\ne \emptyset\}.
\]
We shall denote $C_1(E)$ a set of $1$-chains on the
subset $E$.

Let qubits live on edges $e\in E$.
Accordingly, the Hilbert space is now $(\CC^2)^{\otimes |E|}$,
and basis vectors correspond to $1$-chains $x\in C_1(E)$.
We shall consider a set of $1$-cycles
\be\label{Z(E)}
\calZ(E)=\{ x\in C_1(E)\, : \, \sum_{e\in \delta s\, \cap E} x_e \; \mbox{is even for all $s\in V$}\}.
\ee
For brevity, we shall keep using notation $\calZ\equiv \calZ(E)$ throughout this section.
Define a planar code state $|G_E\ra$ associated with the graph $G_E$ as
\be\label{G_E}
|G_E\ra = \frac1{\sqrt{|\calZ|}}\, \sum_{x\in \calZ} |x\ra.
\ee
Note that $|\calZ|$ can be computed efficiently since $\calZ$ is specified by
mod-$2$ linear equations.

Consider an arbitrary product state
\[
|\Phi\ra = \bigotimes_{e\in E} |\phi_e\ra, \quad |\phi_e\ra= \alpha_e\, |0\ra + \beta_e \, |1\ra.
\]
The quantity we are interested in is the overlap between $|\Phi\ra$ and $|G_E\ra$,
\[
\Gamma = \la G_E|\Phi\ra = |\calZ|^{-\frac12}\, \prod_{e\in E} \alpha_e \sum_{x\in \calZ} \; \prod_{e\in E\, : \, x_e=1}
\left( \frac{\beta_e}{\alpha_e} \right).
\]
This expression becomes singular if $\alpha_e=0$ for some $e$. The singularity can be avoided
by assigning small non-zero value to $\alpha_e$ and using continuity of $\Gamma$
as a function of $\alpha$'s and $\beta$'s (see also~\footnote{One can easily check that
projecting a qubit onto a state $|0\ra$ or $|1\ra$ effectively removes this qubit
from the lattice leading to a state $|G_E\ra$ on a modified planar graph.
One can use this observation to avoid the cases $\alpha_e=0$.}).
For every edge $e\in E$ define a complex weight $w_e=\beta_e/\alpha_e$.
Then $\Gamma$ is proportional to a ``partition function"
\be\label{Z}
Z=\sum_{x\in \calZ}\;  \prod_{e\in E\, : \, x_e=1} w_e.
\ee
%The connection between overlaps of states and partition functions of
%models in statistical mechanics
%is explored in greater generality in \cite{VdN2}.

Let us mention that
$Z$ can be identified with a partition function of
the Ising model with complex weights.  Indeed, introduce virtual Ising spins
$\sigma_f=\pm 1$ living on faces $f$  of $G_E$ and use the ansatz $1-2x_e = \sigma_{f} \sigma_{g}$,
where $f$ and $g$ are the two faces of $G_E$ whose boundary includes the edge $e$.
Then any choice of variables $\sigma_f$ automatically satisfies
the $1$-cycle constraint $x\in \calZ$.
Besides, $w_e^{x_e}=\sqrt{w_e}\, \exp{(-\beta_{f,g}\,\sigma_f \sigma_g)}$, where
$e^{\beta_{f,g}}=\sqrt{w_e}$. Thus $Z$ is proportional to the partition function
of the Ising models with Ising spins living on faces of $G_E$.
In fact, the reverse reduction from the Ising model to the model described by Eq.~(\ref{Z})
is the first step in Barahona's solution of the Ising model~\cite{Barahona81}.
Since some arguments in Barahona's original work  assume that the weights
$w_e$ are real and positive, we shall briefly explain how to apply the same approach to complex weights.
(Note that in general partition functions with complex weights are
more difficult to
compute because of the sign problem.)

\begin{figure}
\centerline{
\mbox{
 \includegraphics[height=3cm]{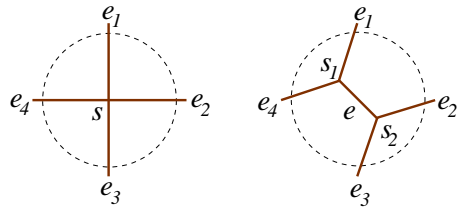}
 }}
\caption{\label{fig:degree4} Local modification of the graph
proposed by Barahona~\cite{Barahona81} to eliminate degree-$4$ vertices.
The two graphs coincide outside the dashed circle. The added edge $e$
carries trivial weight $w_e=1$. The partition functions of the two graphs
coincide.}
\end{figure}

\begin{figure}
\begin{center}
 \includegraphics[height=3.8cm]{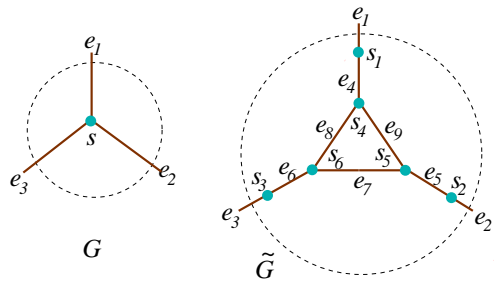}
\end{center}
\caption{\label{fig:degree3} Local modification of the graph
proposed by Barahona~\cite{Barahona81} to reduce summation over $1$-cycles
to summation over perfect matchings. The two graphs coincide outside the dashed circle.
All six added edges $e_4,\ldots,e_9$ carry trivial weight, $w_{e_4}=\ldots=w_{e_9}=1$.
Any $1$-cycle $x\in \calZ$ on the graph $G$ can be uniquely extended to a $1$-chain
$y$ on $\tilde{G}$ such that $x$ and $y$ coincide outside the dashed circle
(including the edges $e_1,e_2,e_3$) and
such that $y$ is a perfect matching inside the dashed circle.
Conversely, $y$ can not be a perfect matching if $y_{e_1}+y_{e_2}+y_{e_3}$
is odd.
}
\end{figure}

Since $G_E$ is a subgraph of the square lattice, vertices of $G_E$ may have degree (number of
incident edges) $1,2,3$, or $4$.
Let us firstly get rid of vertices with degree $1,2,4$.
Suppose $s\in V$ has degree $1$ and $e$ is the edge incident to $s$.
Clearly, $x_e=0$ for any $1$-cycle $x\in \calZ$.
Thus $Z$ does not depend on $w_e$  and we can safely
remove the edge $e$ from the graph.
Now we can assume that all vertices of $G_E$ have degree $2,3$, or $4$.
Suppose $s\in V$ has degree $2$ and $e_1$, $e_2$ are the two edges incident to $s$.
Clearly, $x_{e_1}=x_{e_2}$ for any $1$-cycle $x\in \calZ$.
Therefore we can remove the vertex $s$ and merge the edges $e_1$, $e_2$
into a single edge $e$ with a weight $w_e=w_{e_1} w_{e_2}$.
Now we can assume that all vertices of $G_E$ have degree $3$ or $4$.
Suppose $s\in V$ has degree $4$. Following~\cite{Barahona81} let us
replace $s$ by two vertices $s_1$, $s_2$ of degree $3$
by adding a new edge $e=(s_1,s_2)$ with a weight $w_e=1$, see Figure~\ref{fig:degree4}.
If $e_1,e_2,e_3,e_4$ are the edges incident to $s$, see Figure~\ref{fig:degree4},
and $x\in \calZ$ is a $1$-cycle, then $x_{e_1}+x_{e_2} + x_{e_3} + x_{e_4}=0$
(we use mod-$2$ arithmetics).
After adding a new edge $e$ we get $x_e=x_{e_1}+x_{e_4} =x_{e_2} + x_{e_3}$.
Thus there is one-to-one correspondence between $1$-cycles on the original and
the modified graph. Since $w_e=1$, the partition functions Eq.~(\ref{Z}) are
the same for both graphs. Now we can assume that all vertices of $G_E$ have degree $3$.
Let $G=(V,E)$ be the $3$-valent planar graph that we obtained by
performing all the reductions above (by abuse of notations, we use the
same symbols $V$,$E$ for new set of vertices and edges).
We want to compute the partition function
Eq.~(\ref{Z}) for the graph $G$.
Following~\cite{Barahona81} let us introduce a new planar graph $\tilde{G}=(\tilde{V},\tilde{E})$
such that each vertex $s\in V$
with incident edges $e_1,e_2,e_3\in E$
is replaced by
$6$ vertices  $s_1,\ldots,s_6\in \tilde{V}$ and $6$ edges
$e_4,\ldots,e_9$ as shown on Figure~\ref{fig:degree3}.
We assign trivial weights to all new edges, $w_{e_4}=\ldots=w_{e_9}=1$.
A $1$-chain $y\in C_1(\tilde{E})$ is a perfect matching on the graph $\tilde{G}$ iff
any vertex $s\in \tilde{V}$ has exactly one incident edge $e$ with $y_e=1$.
Let $\calM$ be a set of all perfect matchings on $\tilde{G}$.
As was pointed out by Barahona~\cite{Barahona81}, there is one-to-one
correspondence $\calZ(E)\cong \calM$
between $1$-cycles on $G$ and perfect matching on $\tilde{G}$,
in particular
\be\label{Ztilde}
Z=\sum_{x\in \calZ}\;  \prod_{e\in E\, : \, x_e=1} w_e =
\sum_{y\in \calM}\;  \prod_{e\in \tilde{E}\, : \, y_e=1} w_e.
\ee
\noindent
By construction, $\tilde{G}$ has at least one perfect matching
(the one corresponding to the zero $1$-cycle in $G$), and thus $\tilde{G}$
must have even number of vertices, $|\tilde{V}|=2m$.

Let $A$ be the adjacency  matrix of the graph $\tilde{G}$, i.e.,
$A_{s,t}=1$ if $(s,t)\in \tilde{E}$ and $A_{s,t}=0$ otherwise.
By definition, $A$ is symmetric, $A^T=A$.
One can rewrite the
summation over perfect matchings in
Eq.~(\ref{Ztilde}) as
\be\label{permutations}
Z=\frac1{2^m\, m!}
\sum_{\sigma\in S_{2m}}\,
\prod_{j=1}^m A_{\sigma_{2j-1},\sigma_{2j}}
\,
w_{(\sigma_{2j-1},\sigma_{2j})},
\ee
where we label vertices of $\tilde{G}$ by integers $1,2,\ldots,2m$,
where $\sigma$ is a permutation of the $2m$ vertices,
\[
\sigma =
\left( \ba{ccccccc}
1 & 2 & 3 & 4 & \cdots & 2m-1 & 2m \\
\sigma_1 & \sigma_2 & \sigma_3 & \sigma_4 & \cdots & \sigma_{2m-1} & \sigma_{2m} \\
\ea
\right)
\]
and $S_{2m}$ is the symmetric group.

Kasteleyn has shown~\cite{Kasteleyn61} that
for any planar graph it is possible to find
antisymmetric
matrix $B$
such that
$B_{s,t}=\pm A_{s,t}$ for any pair of vertices $s,t$,
$B^T=-B$, and such that $B$ obeys the {\it Pfaffian orientation}
condition:
\[
\prod_{j=1}^m B_{\sigma_{2j-1},\sigma_{2j}} = \mbox{sgn}\, {(\sigma)}\,
\prod_{j=1}^m A_{\sigma_{2j-1},\sigma_{2j}}.
\]
for any permutation $\sigma\in S_{2m}$.
Here $\mbox{sgn}\, (\sigma)$ is the parity of $\sigma$.
Define a complex antisymmetric $2m\times 2m$ matrix
$B^w$ such that $B^w_{s,t}=B_{s,t} \, w_{(s,t)}$ if $(s,t)\in \tilde{E}$
and $B^w_{s,t}=0$ otherwise.
Then the partition function Eq.~(\ref{permutations}) can be expressed as
the Pfaffian of $B^w$,
\[
Z=\frac1{2^m\, m!} \sum_{\sigma\in S_{2m}} \mbox{sgn}(\sigma)
\,
\prod_{j=1}^m B^w_{\sigma_{2j-1},\sigma_{2j}}=\mbox{Pf}\, (B^w).
\]
Now it can be computed efficiently using the well-known identity
$\mbox{Pf}\, (X)^2 = \mbox{det}\, (X)$ valid for any complex antisymmetric
matrix $X$.

\subsection{Tracing out qubits from the planar code state}
\label{subs:o2}

Suppose at some step $j$ of MQC a subset of qubits
$E=\{e_1,\ldots,e_j\}\subseteq E(\calL)$ has been already  measured.
We want to show that a mixed state
\[
\rho_E=\mbox{Tr}_{e\notin E}\, |K\ra\la K|
\]
describing qubits from $E$ only (before the measurement)
can be represented as a probabilistic mixture of pure states simply related  to
the planar code state  $|G_E\ra$,
see Eq.~(\ref{G_E}).

Indeed, let $\bar{E}=E(\calL)\backslash E$ be the set of traced out qubits.
Denote $\bar{V}\subseteq V(\calL)$ a set of vertices having at least one incident
edge from $\bar{E}$, and $\partial E = V\cap \bar{V}$
be a set of vertices having at least one incident edge from both sets $E$, $\bar{E}$
(it can be regarded as a boundary of the set $E$).
A {\it $0$-chain} on some set of vertices $M$ is
a configuration of binary variables assigned to vertices of $M$,
$u=\{u_s\}_{s\in M}$, $u_s\in \{0,1\}$.
Let us denote $C_0(M)$ a set of all $0$-chains on $M$.
Given any $1$-chain $x\in C_1(E)$, define a {\it boundary} $\partial x \in C_0(V)$
as a $0$-chain such that $(\partial x)_s$ is equal to the parity ($0$ or $1$)
of edges $e$ incident to $s$ with $x_e=1$.
Define a subspace of {\it relative $1$-cycles}
\[
\calZ(E,\partial E)=\{x\in C_1(E)\, : \, (\partial x)_s=0 \; \mbox{for all $s\notin \partial E$}\}.
\]
Thus a relative $1$-cycle satisfies the $1$-cycle condition everywhere except
(may be) of vertices $s\in \partial E$.
Define a subspace of {\it syndromes} as
\[
\calS(E)=\{ u\in C_0(\partial E)\, :\, u=\partial z \; \; \mbox{for some $z\in \calZ(E,\partial E)$}\}.
\]
Simple linear algebra arguments show that
any relative $1$-cycle $z\in \calZ(E,\partial E)$
with a syndrome $u=\partial z\in \calS(E)$
 can be represented as
$z=z(u)+x$, where $z(u)\in \calZ(E,\partial E)$ is some fixed relative $1$-cycle
satisfying $\partial z(u)=u$
and $x\in \calZ(E)$ is $1$-cycle.
Similarly, one can define a subspace of syndroms
\[
\calS(\bar{E})=\{ u\in C_0(\partial {E})\, :\, u=\partial \bar{z} \; \; \mbox{for some $\bar{z}\in
 \calZ(\bar{E},\partial {E})$}\}.
\]
Note that in general $\calS(E)\ne \calS(\bar{E})$.
Any relative $1$-cycle $\bar{z}\in \calZ(\bar{E},\partial E)$
with a syndrome $u=\partial \bar{z} \in \calS(\bar{E})$
can be represented as
$\bar{z}=\bar{z}(u)+y$, where $\bar{z}(u)\in \calZ(\bar{E},\partial E)$ is some fixed relative $1$-cycle
satisfying $\partial \bar{z}(u)=u$
and $y\in \calZ(\bar{E})$ is a $1$-cycle.

The above definitions and arguments imply that
 a Schmidt decomposition of $|K\ra$
with respect to the partition $E(\calL)=E\cup \bar{E}$ can be chosen as follows:
\be\label{Schmidt}
|K\ra=\frac1{\sqrt{|\calS|}}\, \sum_{u\in \calS} |G_E(u)\ra\otimes |G_{\bar{E}}(u)\ra,
\ee
where
\[
\calS=\calS(E)\cap \calS(\bar{E}),
\]
\[
|G_E(u)\ra = \frac1{\sqrt{|\calZ(E)|}}\, \sum_{x\in \calZ(E)}\, |z(u) + x\ra,
\]
and
\[
|G_{\bar{E}}(u)\ra = \frac1{\sqrt{|\calZ(\bar{E})|}}\, \sum_{y\in \calZ(\bar{E})}\, |\bar{z}(u) + y\ra,
\]
Therefore we arrive at
\be
\label{rho_E}
\rho_E=
\frac1{|\calS|}\,
\sum_{u\in \calS}
\,
|G_E(u)\ra\la G_E(u)|.
\ee

\subsection{Calculating probabilities of partial measurements}
\label{subs:o3}

In this section we show how to compute the overlap
$\la \Phi|\rho_E|\Phi\ra$, where $|\Phi\ra$
is an arbitrary product state.
At this point we shall exploit the constraint that $E$ and $\bar{E}$ are connected sets.

The first simplification arising from this constraint is that the subspace of syndromes $\calS$
includes all {\it even} $0$-chains on the boundary $\partial E$:
\be\label{syndromes_are_even}
\calS=\{u\in C_0(\partial E)\, : \, \sum_{s\in \partial E} u_s  = 0 \, \mbox{(mod $2$)}\}.
\ee
Indeed, consider arbitrary even $0$-chain $u\in C_0(\partial E)$, and
let $s_1,s_2,\ldots,s_{2k}\in \partial E$ be the
set of vertices such that $u_{s_j}=1$. Consider firstly the pair of vertices $s_1,s_2$.
Since $s_1,s_2\in \partial E$,
there exist edges $e_1,e_2\in E$ incident to $s_1,s_2$ respectively (may be $e_1=e_2$).
Since $E$ is a connected set, there exists a path $\gamma_{12}\subseteq E$ connecting $e_1$ and $e_2$.
By definition, this path is a relative $1$-cycle, $\gamma_{12}\in \calZ(E,\partial E)$
with the boundary $\partial \gamma_{12}$ supported on $s_1$ and $s_2$.
Similarly, using the connectivity of $\bar{E}$, one can show that there exists
a relative $1$-cycle $\bar{\gamma}_{12}\in \calZ(\bar{E},\partial E)$
with the boundary $\partial \bar{\gamma}_{12}$ supported on $s_1$ and $s_2$.
Repeating this arguments for the remaining pairs of vertices $(s_3,s_4),\ldots, (s_{2k-1},s_{2k})$ we
can construct relative $1$-cycles
$z(u)=\gamma_{12}+\ldots + \gamma_{2k-1,2k}\in \calZ(E,\partial E)$
and $\bar{z}(u)=\bar{\gamma}_{12} + \ldots + \bar{\gamma}_{2k-1,2k}\in \calZ(\bar{E},\partial E)$
such that $\partial z(u)=u$ and $\partial \bar{z}(u)=u$.
Thus $u\in \calS$.
Conversely, for any relative $1$-cycle $z\in \calZ(E,\partial E)$
the boundary $\partial z$ must be even since $z$ consists of elementary edges
and every edge has two endpoints.
The same is true for relative $1$-cycles on $\bar{E}$.
Therefore we have proved Eq.~(\ref{syndromes_are_even}).
It follows that
$|\calS|=2^{|\partial E|-1}$.

The next simplification that we draw from the connectivity of $E,\bar{E}$ is
\begin{lemma}
Suppose $E,\bar{E}$ are connected sets.
Then the graph $G_E$ can be drawn without intersections
on a disk such that all vertices of $\partial E$
lie on the boundary of the disk.
\end{lemma}
{\bf Proof:}
Since $G_E$ is a planar graph,
any point of the plane either belongs to $G_E$, or is `internal' with respect to
$G_E$, or is `external' with respect to $G_E$.
Denote $\mbox{Ext}\, (G_E)$ a set of all external points.
The fact that $E,\bar{E}$ are connected implies that
(1) The boundary
$\partial E$ is contained in the boundary of $\mbox{Ext}\, (G_E)$;
(2) The region $\mbox{Ext}\, (G_E)$ is
topologically equivalent to a plane with a hole.
Therefore one can smoothly deform a plane such that after the deformation
$G_E$ is contained in a disk and
all points of $\partial E$ lie on the boundary of the disk. \qed

This lemma provides us with a way to compute the overlap $\la \Phi|\rho_E|\Phi\ra$.
Indeed, consider two copies of the graph $G_E$ drawn on disjoint disks $D$ and $D^*$.
Let us denote the two copies as $G_E$ and $G_E^*$.
Let us glue the disks $D$, $D^*$ into a sphere such that
vertices of the boundary $\partial E$ on the disk $D$ are glued with the corresponding
vertices of the boundary $\partial E$ on the disk $D^*$.
Denote the resulting graph $G_E\sqcup G_E^*$.
Since $G_E\sqcup G_E^*$ can be drawn on a sphere, it is a planar graph
and thus we can efficiently compute its partition function,
as defined in Eq.~(\ref{Z}).
Let $w_e$ be a weight assigned to edge $e\in E$
on disk $D$.
Let us assign the complex conjugated weight $w_e^*$ to the corresponding edge $e$
on the disk $D^*$.
 For any syndrome $u\in \calS$
let $z(u)\in \calZ(E,\partial E)$
be some fixed relative $1$-cycle such that $\partial z(u)=u$.
Then the  partition function for $G_E\sqcup G_E^*$
can be written as
\[
Z(G_E\sqcup G_E^*) = \sum_{u\in \calS} \, |Z(u)|^2,
\]
where
\[
Z(u)=\sum_{x\in \calZ(E)} \,\;
\prod_{e\in E\, : \, (z(u)+x)_e=1}\, w_e.
\]
One can easily see that $Z(u)$ is equal to
the overlap between a state $|G_E(u)\ra$ in the decomposition Eq.~(\ref{rho_E})
and a product state $|\Phi\ra=\bigotimes_{e\in E} \, (|0\ra + w_e\, |1\ra)$,
that is $Z(u)=\sqrt{|\calZ(E)|}\, \la G_E(u)|\Phi\ra$.
Therefore we conclude that
\[
\la \Phi|\rho_E|\Phi\ra =
\frac1{2^{|\partial E|-1}\, |\calZ(E)|}\,
Z(G_E\sqcup G_E^*)
\]
and thus the overlap can be computed efficiently
using techniques from the previous section.

We can use the algorithm above for computing the overlap between
$\rho_E$ and a product state to compute
probabilities of any particular partial measurement outcomes $p(m_1,\ldots,m_j)$.
It allows us to compute {\it conditional probabilities}
$p(m_j \, || \, m_1,\ldots,m_{j-1})$ for any $j=1,\ldots,N_1$.
If we already know the outcomes $m_1,\ldots,m_j$, we can
sample $m_j$ from the probability distribution
$p(m_j \, || \, m_1,\ldots,m_{j-1})$.
Thus all quantum steps in MQC can be efficiently simulated on a
classical computer (with random numbers).

\section{Extension to non-planar graphs}

In this section we outline an MQC simulation scheme built around the
planarity of graphs. It is complementary to the simulation scheme
\cite{Shi1} - \cite{VandenNest06} via tensor networks,
centered around the tree-ness of graphs. We identify a parameter
$\eta$ which measures deviation from planarity and show that
simulation of MQC is efficient in the number of particles in the
resource state but exponentially inefficient in $\eta$.

In Section~\ref{noTTN} it is shown that simulation based on
planarity can not be subsumed under the tree tensor network method.
In Section~\ref{uC} we show that there is another interesting
example among the states for which MQC can be related to the Ising
model partition function: a universal 2D cluster state. The
corresponding interaction graph (to be distinguished from the graph
upon which the definition of graph states is based) is highly
non-planar, such that simulation---as expected---is not efficient.
In Section~\ref{switchM} we characterize the complexity of classical
MQC simulation for non-planar Ising interaction graphs.

We would like to remark that the connection between the 1D cluster
state and non-interacting fermions, a topic interwoven with the
solvability of the Ising model on planar graphs, has previously been
discussed in \cite{Pachos}.

\subsection{The planar code state and tensor contraction}
\label{noTTN}

Here we show that a planar code state on a lattice of size
$L \times L$ has an {\em{entanglement width}} of at least $L$ such
that its simulation
scheme based on a TTN is exponentially inefficient in $L$. See \cite{VdN2}
for an alternative proof of this result.

The entanglement measure entanglement width has been defined in
\cite{VdN1} and related to universality of MQC and to the complexity
of its TTN-simulation in \cite{VdN1, VandenNest06}. We briefly restate
the definition here. Consider a quantum system on a set $Q$ of qubits. Be $T$ a
tree graph with each vertex having 1 or 3 incident edges (subcubic tree). The
vertices with one edge are called leaves, and each of them corresponds
to a qubit of the considered quantum system. If an edge $e$ is deleted
from $T$, the resulting graph $T\backslash e$ has two connected
components which induce a bi-partition $(A_T^e, B_T^e)$ of $Q$. For a
pure state $|\psi\rangle$ we denote
by $E_{A_T^e,B_T^e}(|\psi\rangle)$ the entanglement entropy
with respect to that bi-partition. The entanglement width is now
defined as
\[
E_{\text{wd}}(|\psi\rangle):= \min_T \max_e E_{A_T^e,B_T^e}(|\psi\rangle),
\]
where the minimization is over all subcubic trees with $n$ leaves, and
the maximum over all edges $e$ in a given tree.

\begin{lemma}
  The entanglement width of a planar code state $|G_L\rangle$ on a
  square lattice of size $L \times L$ is
  \be
  E_{\text{\em{wd}}}(|G_L\rangle) \geq L.
  \ee
\end{lemma}
\textbf{Proof:} First we show that for any given subcubic tree $T$ with $|Q|$
leaves, there exists
an edge $e$ such that $1/3\, |Q| \leq |A_T^e|, |B_T^e| \leq 2/3\,
|Q|$. Given $T$, pick an arbitrary 3-valent
vertex as root. This induces an ancestor-descendant relation among pairs
of vertices. For each vertex $a \in V(T)$, denote by $n(a)$ the number
of leaves among $a$ and all its descendants. There are vertices $a \in
V(T)$ for which $n(a)>2/3 n$. Among those, there must be a vertex
$a_0$ such that for both its direct descendants $a_1, a_2$ holds $n(a_1),
n(a_2)\leq 2/3\, n$. Choose $a_1$ s.th. $n(a_1) \geq n(a_2)$. The
desired edge then is $e=(a_0,a_1)$. Be $A_T^e$ the set of qubits
associated with the descendant leaves of $a_1$. Then, $|A_T^e|=n(a_1)\leq
2/3\, n$. Also, $1/3\, n < n(a_0)/2 = (n(a_1)+n(a_2))/2 \leq n(a_1)
=|A_T^e|$.

Now, the entanglement entropy $E_{A_T^e,B_T^e}$ is linear in the
length of the boundary between $A_T^e$ and $B_T^e$ \cite{Zanardi04}.
It is minimized if $A_T^e$ and $B_T^e$ are simply connected and use
the external boundary of the $L \times L$ lattice. This occurs e.g.
for $A_T^e$ filling up the bottom part of the lattice up to a
certain height. The entanglement entropy in that case is $L$ or
$L+1$. Thus, $\max_e E_{A_T^e,B_T^e}(|G_L\rangle) \geq L$, for all
subcubic trees $T$. \qed

The planar code state
considered here is, like any other stabilizer state, local Clifford
equivalent to a graph state \cite{Schling, Grassl, Hein}. For graph states, the
entanglement width equals the so-called rank width \cite{Oum} of the
underlying graph. The rank width $\chi$ is the critical parameter for the
complexity of MQC simulation via tree tensor networks; the operational
resources required in the simulation of an $n$-qubit system scale like
$Poly(n, 2^\chi)$
\cite{VandenNest06} (c.f. Theorem 4 therein). Thus, MQC on $L \times
L$-planar code states, with $\chi$ growing at least linearly in $L$,
is not suited for efficient simulation by the methods developed in
\cite{Shi1} - \cite{VandenNest06}.

\subsection{The 2D cluster state in the Ising model}
\label{uC}

\begin{figure}
  \begin{center}
    \includegraphics[width=5.3cm]{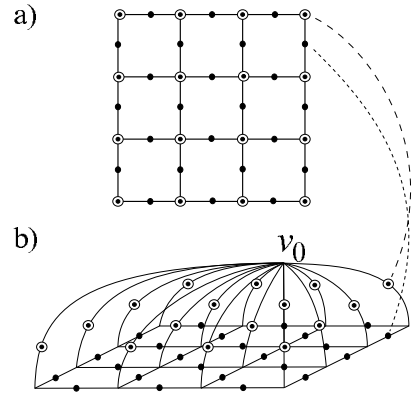}
  \end{center}
  \caption{\label{non_planar}Universal graph state and its surface code
    representation. a) The two-dimensional cluster state
    $|{\cal{C}}_L\rangle$, for $L=3$. Qubits
    are located on the vertices $\bullet$, $\odot$. b) Local unitary
    equivalent surface code state $|\tilde{G}_L\rangle$. Here, qubits are
    located on the edges, and state stabilizers are associated with
    the sites and the plaquettes. To better illustrate the identification of
    qubits in $|\tilde{G}_L\rangle$ with their counterparts in
    $|{\cal{C}}_L\rangle$, the edges of $\tilde{G}_L$ are labeled by the
    same symbols as the vertices
    in ${\cal{C}}_L$. The dashed lines indicate pairs of corresponding qubits.}
\end{figure}

There is a two-dimensional universal
cluster state associated with an edge-centered graph (see
Fig.~\ref{non_planar}a) whose overlap with
an arbitrary product state is
described by the Ising model partition function. However, its interaction
graph is not planar.

The cluster state $|{\cal{C}}_L\rangle$ under consideration is shown in
Fig.~\ref{non_planar} a. Qubits are
associated with the vertices of the displayed graph ${\cal{C}}_L$. The
stabilizer of
${\cal{C}}_L$ is generated by the Pauli operators
\[
K_a = X_a \bigotimes_{b|\, (a,b)
  \in E({\cal{C}})} Z_b,\;\; \forall a \in V({\cal{C}}_L).
\]
The graph ${\cal{C}}_L$ is bi-colorable. We refer to the two colors as
``$R$'' and ``$B$'', and use symbols ``$\odot$'' and
``$\bullet$'' to display the respective vertices. The cluster state
$|{\cal{C}}_L\rangle$ is a {\em{universal}}
resource for quantum computation \cite{Debbie}.

Fig.~\ref{non_planar}b shows a surface code state $|\tilde{G}_L\rangle$
based on non-planar graph $\tilde{G}_L$. Qubits are associated with
edges in $\tilde{G}_L$. $\tilde{G}_L$ is a modification of the previously
discussed $L\times L$-lattice graph. It has one extra vertex $v_0$,
extra edges $(v_0,v_i)$ and
extra plaqettes $(v_0,v_i,v_j)$, where $v_i,v_j$ are vertices in the
plane incident to a common edge. The stabilizer of
$|\tilde{G}_L\rangle$ is spanned by the operators $A_s$, $B_p$ defined
in Eq. (\ref{SG}). For an $L \times L$-lattice there are $N_1^\prime =
2L(L+1)+(L+1)^2$ qubits, $N_0^\prime=(L+1)^2$ independent site-
and $N_2^\prime = 2L(L+1)$ independent plaquette stabilizers, and
$|\tilde{G}_L$ is thus uniquely specified. $|\tilde{G}_L\rangle$ may
again be written in the form of Eq. (\ref{G_E}), and its overlap with
a multi-local state may
again, as in Eq. (\ref{Z}) be related to the partition function of the
Ising model. However, the underlying interaction graph $\tilde{G}_L$ is now
non-planar such that the techniques for efficient simulation
\cite{Kasteleyn61,Barahona81} are no longer applicable.

The cluster state $|{\cal{C}}_L\rangle$ and surface code state
$|\tilde{G}_L\rangle$ are local unitary equivalent. First, they have
support on the same number of qubits, and we can pairwise identify
the respective qubit locations. The mapping is as follows. The
vertices of color $R$ in ${\cal{C}}$ reappear as vertices of
$\tilde{G}_L$, but they are no longer qubit locations. As for the
latter, each $R$-colored vertex $v$ of ${\cal{C}}_L$ corresponds to
an edge $(v,v_0)$ of $\tilde{G}_L$ above the plane, and each
$B$-colored vertex $v^\prime$ of ${\cal{C}}_L$, located between two
$R$-colored vertices $v_1$, $v_2$, corresponds to the edge
$(v_1,v_2)$ of $\tilde{G}_L$ inside the plane; See
Fig.~\ref{non_planar}. Denote by $H_R$ the simultaneous local
Hadamard transformation on all qubits of color $R$. Then,
$|\tilde{G}_L\rangle = H_R |\phi\rangle_{{\cal{C}}_L}$.

To see this, first consider the stabilizer $K_w$ of
$|{\cal{C}}_L\rangle$ associated with a
$B$-colored qubit location $w$, $K_w = X_w Z_{v_1} Z_{v_2}$ ($v_1,\, v_2$
are vertices of color $R$). Under the identification of vertices
in ${\cal{C}}_L$ with edges in $\tilde{G}_L$, $v_{1,2}
\leftrightarrow (v_0,v_{1,2})$, $w \leftrightarrow (v_1,v_2)$. (Again,
$v_0$ is the extra vertex above the code plane.) After
application of $H_R$, the operator $K_w$ is mapped to
$X_{(v_1,v_2)}X_{(v_0,v_1)}X_{(v_0,v_2)} = B_p$ for $p =
(v_0,v_1,v_2)$; c.f. Eq. (\ref{SG}). Second, consider the stabilizer
$K_v$ for an $R$-colored vertex $v \in V({\cal{C}}_L)$, $K_v=X_v
\otimes_{w|\,(v,w)
  \in E({\cal{C}})}Z_w$. With the identification $v \leftrightarrow
  (v,v_0)$, $w \leftrightarrow (v, w)$, and after application of
  $H_R$, the stabilizer $K_v$ is mapped onto $Z_{(v,v_0)}
  \otimes_{w|(v,w) \in E(\tilde{G}_L)}Z_{(v,w)} = A_v$. \qed

The overlap between a local state $|\Phi\rangle =
\bigotimes_{e \in \tilde{E}}|\phi_e\rangle$ and $|\tilde{G}_L\rangle$
may again be written as the partition function of the Ising
model,
\begin{equation}
  \label{IsingMF}\
  \begin{array}{rcl}
    \displaystyle{\langle \tilde{G}_L|\Phi \rangle} &\sim&
    \displaystyle{\sum_{\{\sigma_j\}}
    \exp\left(\sum_{(j,k)}\beta_{jk}
      \sigma_j \sigma_k \right)}\\
    &\sim& \displaystyle{\sum_{\{\sigma_j| j \neq v_0\}}
    \!\!\!\!\exp\left(\sum_{(jk)|j,k \neq v_0} \!\!\!\!\! \beta_{jk}
      \sigma_j \sigma_k + \sum_{j\neq v_0} \!\! \beta_{j0}\sigma_j\right).}
\end{array}
\end{equation}
Therein, the $\beta_{jk}$ are specified through the relation
$\exp(\beta_{jk}) = \sqrt{\langle 0| \tilde{\phi}_{jk}\rangle/\langle
  1 | \tilde{\phi}_{jk}\rangle}$,
and $|\tilde{\phi}_e\ra=H_e\, |\phi_e\ra$.
   As can be seen from the lower line in
Eq. (\ref{IsingMF}), the overlap between a local state and a
universal 2D cluster state corresponds to the partition function of
the planar Ising
model in the presence of a magnetic field.

\subsection{The complexity of MQC simulation for non-planar
  interaction graphs}
\label{switchM}

\begin{figure}
  \begin{center}
    \includegraphics[width=5.5cm]{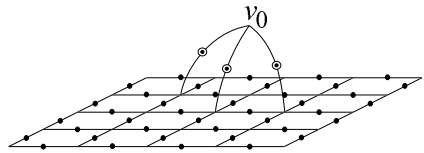}
  \end{center}
  \caption{\label{intermediate}Intermediate state between the planar
    code state and a two-dimensional cluster state.}
\end{figure}

First, we would like to morph the simulation of MQC with a planar
code state into simulation of MQC with the universal cluster state,
by `switching on magnetic fields'. This may be performed gradually,
switching on one magnetic field at a time. Fig.~\ref{intermediate}
shows the graph corresponding to an intermediate state of this
sequence. The sequence is passed in reverse order if one starts from
a 2D cluster state of Fig.~\ref{non_planar} and measures, one by
one, the cluster qubits of color $R$ in the $X$-basis. More
generally, $X$- and $Z$-measurements directly operate at the level
of Ising interaction graphs $\tilde{G}$; by deletion and contraction
of edges, respectively.

A measure for deviation from a planar graph is the
number $\eta$ of edges that needs to be deleted from a graph to make it
planar. This
parameter $\eta$ also governs the complexity of a straightforward
extension of the simulation scheme presented in Section~\ref{CS}, applicable to
non-planar graphs. Be $E_\text{np}$ a set of edges
in $\tilde{G}$ such that $\tilde{G}\backslash E_\text{np}$ is
planar. Then, the product state $\bigotimes_{e \in
  E_\text{np}} |\phi_{e}\rangle$ may be expanded into the $X$-basis and the
efficient simulation of MQC on the remaining planar graph may be run for each
component. The operational cost of this simulation method for an
$n$-qubit state $|\tilde{G}\rangle$ is $Poly(n)2^\eta$, which may be
compared to Theorem 4 of \cite{VandenNest06} for tree tensor networks.

\section{Conclusion}

We have shown that the overlap between the planar code
state and any product state can be computed efficiently
classically, through its
correspondence with the planar Ising model. Under rather general
assumptions about
the pattern of one-qubit measurements, MQC
with the  planar code state as a quantum
resource is not universal.
It can be efficiently simulated on a classical computer.
MQC with a universal 2D cluster state can also be related to the Ising
model. However, the corresponding interaction graph is non-planar.

\acknowledgements{We are thankful to Alexei Kitaev for
useful discussions and reading the manuscript.
RR would also like to thank Panos Aliferis, Maarten van
  den Nest and Herbert Wagner for discussions. SB acknowledges support
  by the NSA and ARDA through ARO contract number W911NF-04-C-0098.
RR
is supported by the Government of Canada through NSERC and by the
  Province of Ontario through MEDT.}

\end{document}